\documentclass{article}
\usepackage{spconf,amsmath,graphicx,hyperref}
\usepackage{cite}
\usepackage{amsmath,amssymb,amsfonts}
\usepackage{graphicx}
\usepackage{textcomp}
\usepackage{amsmath}
\usepackage{booktabs} 
\usepackage{multirow}
\usepackage{subcaption}
\usepackage{hyperref}
\usepackage{tabularx}
\usepackage{url}            
\usepackage{booktabs}       
\usepackage{amsfonts}       
\usepackage{nicefrac}       
\usepackage{microtype}      
\usepackage{xcolor}         
\usepackage{algorithm}
\usepackage{colortbl}
\usepackage{algpseudocode}
\usepackage{amsmath}
\usepackage{tikz}
\usepackage{orcidlink}
\usepackage[font=small]{caption}
\usepackage{float}
\usepackage{url}
\usepackage{stfloats}

\usepackage{orcidlink}
\usepackage{comment}
\usepackage{amssymb}
\usepackage{graphicx}
\usepackage{braket}
\usepackage[table]{xcolor}
\usepackage[skip=0pt,font=small]{caption}
\usepackage{booktabs}
\usepackage{setspace}
\usepackage{wrapfig}
\usepackage{tabularx}\definecolor{classicalrow}{RGB}{245,248,255}

\definecolor{qtlrow}{RGB}{245,239,252} 

\definecolor{ketgptrow}{RGB}{245,239,252}
\definecolor{fixedrow}{RGB}{245,239,252}
\definecolor{lowconfrow}{RGB}{213, 223, 247}
\definecolor{adaptiveqrow}{RGB}{213, 223, 247} 
\definecolor{candidate160row}{RGB}{226,244,226} 
\usepackage{xcolor}

\usepackage{stfloats}
\title{
QSTAR: Quantum Selective Transfer with Adaptive Routing
}
\name{
Saim Rehman,
Nouhaila Innan,
Muhammad Shafique
}

\address{
\small eBRAIN Lab, Division of Engineering, New York University Abu Dhabi (NYUAD), Abu Dhabi, UAE\\
\small Center for Quantum and Topological Systems (CQTS), NYUAD Research Institute, NYUAD, Abu Dhabi, UAE\\
\small \{sr7849, nouhaila.innan, muhammad.shafique\}@nyu.edu
}
\begin{document}
%
\maketitle
\begin{abstract}
Quantum transfer learning (QTL) is commonly evaluated by replacing a classical classifier with a fixed variational quantum head, leaving unclear when quantum inference provides measurable benefit. We propose \textbf{QSTAR: Quantum Selective Transfer with Adaptive Routing}, which retains high-confidence classical predictions and routes low-confidence samples to a quantum fallback. Using a frozen ResNet18 backbone, we compare manually designed QTL heads, architecture-searched KetGPT heads, and parameter-matched classical baselines under consistent experimental settings. On Fashion-MNIST, standard QTL heads reach at most $59.7\%$ accuracy and $0.564$ F1-score, whereas compact KetGPT Candidate \#160 reaches $80.2\%$ accuracy and $0.800$ F1-score. On the routed low-confidence subset, Candidate \#160 improves accuracy over the matched MLP by $6.31$, $5.38$, and $3.73$ percentage points at $\tau=0.70$, $0.80$, and $0.90$, respectively. At the selected operating point $\tau=0.70$, the adaptive system reaches $81.78\%$ accuracy and $0.816$ F1-score while routing only $36.12\%$ of test samples to the quantum branch. Candidate \#160 uses 10 quantum parameters and 9 gates, while experiments on CIFAR-10, KMNIST, and SVHN show consistent accuracy gains over the corresponding adaptive classical baselines. These results show that selective quantum invocation can reduce quantum execution while improving inference on low-confidence samples.
\end{abstract}
\section{Introduction}

Quantum transfer learning (QTL) combines pretrained classical feature extraction with a variational quantum classifier, reducing the dimensionality presented to the quantum model and making hybrid learning suitable for near-term devices~\cite{mari2020transfer}. Most QTL studies, however, use the quantum head for every sample, making performance highly dependent on circuit design, qubit count, depth, encoding, and measurement budget~\cite{kim2023classical,mogalapalli2022classical,rehman2026fairbenchmark}. This also makes it difficult to isolate the contribution of the quantum component from the backbone, compression, optimization, and parameter count.

We ask a different question: \emph{when should the quantum branch be invoked?} We propose \textbf{QSTAR: Quantum Selective Transfer with Adaptive Routing}, which retains high-confidence classical predictions and routes low-confidence samples to a quantum fallback following selective-classification principles~\cite{geifman2017selectiveclassificationdeepneural}. QSTAR uses a frozen ResNet18 backbone and compares manually designed QTL heads with architecture-searched KetGPT circuits~\cite{apak2024ketgpt}. The selected quantum fallback is further compared with an MLP with approximately matched total trainable parameter count under the same routing protocol.

\textbf{The main contributions are:}
\textbf{(i)} a selective QTL framework that activates the quantum branch only for low-confidence samples;
\textbf{(ii)} a controlled comparison of manually designed and KetGPT-searched quantum heads against classical baselines;
\textbf{(iii)} an adaptive evaluation showing that Candidate \#160 improves low-confidence accuracy over the matched MLP across all tested thresholds and reaches $81.78\%$ adaptive accuracy; and
\textbf{(iv)} a resource study showing that Candidate \#160 requires only 10 quantum parameters and 9 gates while limiting quantum execution to a subset of test samples.

\section{Methodology}

\begin{figure*}[htpb]
    \centering
    \includegraphics[width=\linewidth]{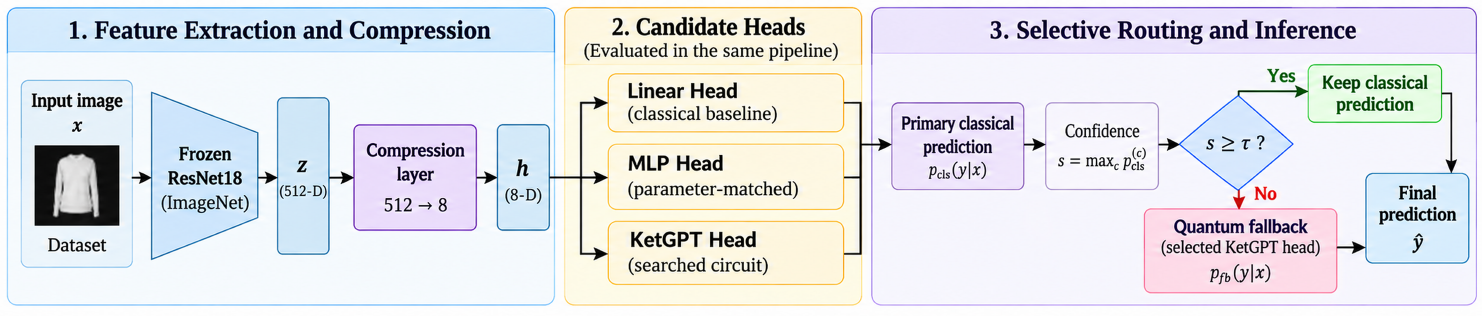}
    \caption{Overview of QSTAR. A frozen ResNet18 extracts image features that are compressed and evaluated by classical and quantum candidate heads. During adaptive inference, the primary classical prediction is retained when its confidence exceeds threshold $\tau$; otherwise, the sample is routed to the selected KetGPT quantum fallback.}
    \label{fig:qstar_method0}
\end{figure*}

QSTAR combines transfer learning, searched quantum classifier design, and confidence-based adaptive routing in a single framework. A frozen ResNet18 backbone extracts image features, which are compressed before classification by classical or quantum heads. During inference, the primary classical classifier also provides a confidence score: confident samples retain the classical prediction, while uncertain samples are sent to the selected KetGPT quantum fallback. The full workflow is illustrated in Fig.~\ref{fig:qstar_method0}.

\subsection{Transfer Learning and Candidate Heads}

For input image $x_i$, where $i$ indexes the samples, QSTAR uses a frozen ImageNet-pretrained ResNet18 backbone to extract the $512$-dimensional feature representation
$z_i=f_{\theta_0}(x_i)\in\mathbb{R}^{512}$,
where $f_{\theta_0}$ denotes the feature extractor and $\theta_0$ its fixed parameters~\cite{he2015deepresiduallearningimage}. A trainable compression function $\phi_{\omega}$, parameterized by $\omega$, maps this representation to
$h_i=\phi_{\omega}(z_i)\in\mathbb{R}^{d}$,
where $d$ is the compressed feature dimension and $d=8$ in the main experiments.
We evaluate a linear classical head, an MLP with approximately matched total trainable parameter count, manually designed QTL heads, and KetGPT-generated quantum heads using the same feature pipeline. For a quantum head, $h_i$ is encoded into a $q$-qubit parameterized quantum state $|\psi(h_i,\alpha)\rangle=
U_{\mathrm{var}}(\alpha)
U_{\mathrm{enc}}(h_i)
|0\rangle^{\otimes q},$
where $q$ is the number of qubits, $U_{\mathrm{enc}}$ is the feature-encoding unitary, $U_{\mathrm{var}}$ is the variational circuit, and $\alpha$ denotes its trainable quantum parameters. For observable $M_j$, indexed by $j$, the circuit produces the expectation value $m_j(h_i,\alpha)=
\langle\psi(h_i,\alpha)|M_j|\psi(h_i,\alpha)\rangle.$
The resulting measurement vector $m(h_i,\alpha)$ is mapped to the $C$ class logits as
$\ell_i=W_qm(h_i,\alpha)+b_q$,
where $C$ is the number of classes and $W_q$ and $b_q$ are the trainable weights and bias of the classical readout.
Standard QTL heads use manually specified circuit templates with different qubit counts and depths. KetGPT heads use searched circuit structures, including their gate sequence, parameter placement, and entangling topology~\cite{apak2024ketgpt}.

\subsection{KetGPT Candidate Selection}

Each KetGPT candidate, indexed by $k$, is characterized by its qubit count $q_k$, number of trainable quantum parameters $P_k$, and gate count $G_k$. The main search retains 8-qubit circuits satisfying
$10\leq P_k\leq80$ and $G_k\leq150$.
Candidates are checked for valid circuit construction and finite measurement outputs before training.
Each accepted circuit topology $U_k(\alpha_k)$ is evaluated using the same transfer-learning pipeline: $|\psi_k(h_i,\alpha_k)\rangle=
U_k(\alpha_k)
U_{\mathrm{enc}}(h_i)
|0\rangle^{\otimes q_k},$
where $\alpha_k$ denotes the trainable parameters of candidate $k$. The topology remains fixed while $\alpha_k$, the compression parameters $\omega$, and the classical readout parameters are optimized. Candidate \#160 is additionally evaluated in a compact-circuit study in which the lower bound on $P_k$ is removed, allowing circuits with fewer than 10 trainable quantum parameters.

\subsection{Selective Routing and Resource-Aware Inference}

QSTAR activates the quantum branch only when the primary classical classifier is uncertain. Let $p_{\mathrm{cls},i}^{(c)}$ denote the probability assigned by the primary classifier to class $c\in\{1,\ldots,C\}$ for sample $i$. Its confidence score is
$s_i=\max_c p_{\mathrm{cls},i}^{(c)}$.

Given confidence threshold $\tau$, the routing decision is
\begin{equation}
r_i=
\begin{cases}
0, & s_i \geq \tau,\\
1, & s_i < \tau,
\end{cases}
\end{equation}
where $r_i=0$ retains the primary classical prediction and $r_i=1$ sends the sample to the fallback head. The final prediction is
\begin{equation}
\hat{y}_i=
\begin{cases}
\arg\max_c p_{\mathrm{cls},i}^{(c)}, & r_i=0,\\
\arg\max_c p_{\mathrm{fb},i}^{(c)}, & r_i=1,
\end{cases}
\end{equation}
where $p_{\mathrm{fb},i}^{(c)}$ denotes the probability assigned to class $c$ by either the matched MLP or the selected KetGPT fallback.

For a test set of $N_{\mathrm{test}}$ samples, the routed fraction is $R(\tau)=
\frac{1}{N_{\mathrm{test}}}
\sum_{i=1}^{N_{\mathrm{test}}}r_i.$
If each routed quantum inference uses $S$ measurement shots, the average quantum shot cost per test sample is
$\bar{S}(\tau)=R(\tau)S$.
Quantum execution is thus required only for the fraction $R(\tau)$ of samples selected by the confidence router.

The operating threshold $\tau=0.70$ is selected using validation performance, while $\tau\in\{0.70,0.80,0.90\}$ is reported to characterize the accuracy--resource trade-off. Images are resized to $224\times224$, converted to three channels, and normalized using ImageNet statistics~\cite{xiao2017fashion}. Models are trained for 10 epochs using Adam and cross-entropy loss. Quantum circuits are simulated in PennyLane with 100 measurement shots and parameter-shift gradients~\cite{Wierichs_2022,Schuld_2019}. The learning rate, batch size, weight decay, simulator backend, and initialization settings are kept fixed within each experiment family.
\section{Results \& Evaluation}

All implementation details, hyperparameter settings, experimental configurations, and result files are provided in the accompanying code repository to support reproducibility\footnote{\url{https://github.com/eBrain4Everyone/QSTAR}}. QSTAR is evaluated on four image-classification datasets. We provide a detailed experimental analysis on Fashion-MNIST, covering fixed-head performance, low-confidence fallback behavior, adaptive routing, and resource characteristics. Results for the remaining datasets are summarized through a cross-dataset study to assess whether the observed behavior remains consistent across different data distributions. The KetGPT circuits used in our experiments are loaded automatically through PennyLane, while the original circuit library is publicly available in the KetGPT dataset\footnote{\url{https://www.kaggle.com/datasets/boranapak/ketgpt-data}}.

\subsection{Fixed-Head Transfer Learning}

As shown in Table~\ref{tab:qtl_baselines_fmnist}, we first evaluate the classifier heads in a fixed transfer-learning setting, where every sample is processed by the same head. The manually designed QTL models exhibit strong sensitivity to circuit configuration, with accuracy ranging from $39.3\%$ to $59.7\%$ across the evaluated qubit counts and circuit depths. Increasing the number of qubits or circuit depth does not consistently improve performance; for example, the 8-qubit depth-1 model reaches $44.6\%$, while the 6-qubit depth-4 model reaches $59.7\%$.
The searched KetGPT heads achieve substantially stronger fixed-head performance. Compact Candidate \#160 reaches $80.2\%$ accuracy and an F1-score of $0.800$, approaching the strongest classical linear head at $80.8\%$ accuracy while using only 10 trainable quantum parameters. This motivates its use as the quantum fallback in the adaptive QSTAR evaluation.

\begin{table*}[!t]
\centering
\caption{Fixed-head comparison on Fashion-MNIST using a frozen ResNet18 backbone.
Results are mean $\pm$ sample standard deviation over five seeds; quantum models use 100 shots.}
\label{tab:qtl_baselines_fmnist}
\small
\setlength{\tabcolsep}{5pt}
\renewcommand{\arraystretch}{1.05}
\resizebox{\textwidth}{!}{%
\begin{tabular}{llccccccc}
\toprule
\textbf{Model} &
\textbf{Head} &
\textbf{Qub.} &
\textbf{Depth} &
\textbf{Head Params.} &
\textbf{Acc. (\%)} &
\textbf{F1} &
\textbf{AUC} &
\textbf{Train Time} \\
\midrule

\rowcolor{classicalrow}
\textbf{Classical}
& Linear & -- & -- & 90
& $\mathbf{80.8 \pm 0.8}$
& $\mathbf{0.807 \pm 0.010}$
& $\mathbf{0.977 \pm 0.002}$
& $1.08 \pm 0.01$ h \\

& MLP & -- & -- & 162
& $80.0 \pm 1.1$
& $0.800 \pm 0.013$
& $0.974 \pm 0.003$
& $1.08 \pm 0.01$ h \\

\midrule

\textbf{Standard QTL}
& Var. & 4 & 1 & 12
& $39.3 \pm 7.7$
& $0.305 \pm 0.089$
& $0.892 \pm 0.020$
& $2.78 \pm 0.02$ h \\

& Var. & 4 & 2 & 24
& $44.3 \pm 2.9$
& $0.370 \pm 0.038$
& $0.898 \pm 0.013$
& $5.17 \pm 0.05$ h \\

& Var. & 4 & 4 & 48
& $45.3 \pm 5.6$
& $0.387 \pm 0.065$
& $0.894 \pm 0.025$
& $13.01 \pm 0.09$ h \\

& Var. & 6 & 1 & 18
& $41.5 \pm 10.4$
& $0.323 \pm 0.109$
& $0.884 \pm 0.048$
& $4.59 \pm 0.04$ h \\

& Var. & 6 & 2 & 36
& $50.2 \pm 5.8$
& $0.433 \pm 0.078$
& $0.906 \pm 0.014$
& $9.79 \pm 0.05$ h \\

& Var. & 6 & 4 & 72
& $\mathbf{59.7 \pm 5.6}$
& $\mathbf{0.564 \pm 0.069}$
& $\mathbf{0.927 \pm 0.013}$
& $27.15 \pm 0.25$ h \\

& Var. & 8 & 1 & 24
& $44.6 \pm 4.9$
& $0.350 \pm 0.051$
& $0.898 \pm 0.024$
& $7.06 \pm 0.09$ h \\

& Var. & 8 & 2 & 48
& $51.2 \pm 10.0$
& $0.447 \pm 0.132$
& $0.911 \pm 0.031$
& $16.37 \pm 0.18$ h \\

\midrule

\rowcolor{candidate160row}
\textbf{KetGPT}
& Cand. \#160 & 8 & -- & 10
& $\mathbf{80.2 \pm 1.2}$
& $\mathbf{0.800 \pm 0.013}$
& $\mathbf{0.975 \pm 0.001}$
& $\mathbf{3.74 \pm 0.03}$ h \\

& Cand. \#180 & 8 & -- & 58
& $76.7 \pm 2.0$
& $0.756 \pm 0.027$
& $0.966 \pm 0.007$
& $22.90 \pm 0.34$ h \\

\bottomrule
\end{tabular}%
}
\end{table*}

\begin{table*}[ht]
\centering
\caption{Fashion-MNIST routing results using compact KetGPT Candidate \#160 as the quantum fallback.
Results are mean $\pm$ sample standard deviation over five seeds.
Routed denotes the percentage of test samples sent to the fallback branch;
Params denotes total trainable parameters.}
\label{tab:adaptive_ketgpt}
\small
\setlength{\tabcolsep}{4pt}
\renewcommand{\arraystretch}{1.08}
\resizebox{\textwidth}{!}{%
\begin{tabular}{lcccccccc}
\toprule
\textbf{Model} &
\textbf{$\tau$} &
\textbf{Routed (\%)} &
\textbf{Params} &
\textbf{Acc. (\%)} &
\textbf{Prec.} &
\textbf{Rec.} &
\textbf{F1} &
\textbf{AUC} \\
\midrule

\multicolumn{9}{c}{\textbf{Low-Confidence Subset}} \\
\midrule

MLP & 0.70
& $34.9 \pm 3.8$ & 4266
& $53.97 \pm 4.32$
& $0.559 \pm 0.043$
& $0.546 \pm 0.032$
& $0.525 \pm 0.030$
& $0.902 \pm 0.020$ \\

\rowcolor{candidate160row}
KetGPT \#160 & 0.70
& $36.12 \pm 3.86$ & 4204
& $\mathbf{60.28 \pm 1.32}$
& $\mathbf{0.615 \pm 0.063}$
& $\mathbf{0.593 \pm 0.035}$
& $\mathbf{0.584 \pm 0.021}$
& $\mathbf{0.927 \pm 0.007}$ \\

\addlinespace[2pt]

MLP & 0.80
& $50.4 \pm 4.8$ & 4266
& $61.70 \pm 4.56$
& $0.633 \pm 0.054$
& $0.616 \pm 0.027$
& $0.600 \pm 0.037$
& $0.924 \pm 0.012$ \\

\rowcolor{candidate160row}
KetGPT \#160 & 0.80
& $50.88 \pm 2.66$ & 4204
& $\mathbf{67.08 \pm 2.17}$
& $\mathbf{0.676 \pm 0.044}$
& $\mathbf{0.657 \pm 0.022}$
& $\mathbf{0.653 \pm 0.013}$
& $\mathbf{0.945 \pm 0.007}$ \\

\addlinespace[2pt]

MLP & 0.90
& $67.9 \pm 4.0$ & 4266
& $69.03 \pm 3.21$
& $0.710 \pm 0.039$
& $0.696 \pm 0.020$
& $0.688 \pm 0.031$
& $0.945 \pm 0.009$ \\

\rowcolor{candidate160row}
KetGPT \#160 & 0.90
& $66.34 \pm 2.61$ & 4204
& $\mathbf{72.76 \pm 1.76}$
& $\mathbf{0.751 \pm 0.034}$
& $\mathbf{0.731 \pm 0.012}$
& $\mathbf{0.733 \pm 0.016}$
& $\mathbf{0.959 \pm 0.005}$ \\

\midrule

\multicolumn{9}{c}{\textbf{Full Adaptive System}} \\
\midrule

Classical & 0.70
& $34.9 \pm 3.8$ & 8460
& $80.02 \pm 1.27$
& $0.802 \pm 0.020$
& $0.802 \pm 0.013$
& $0.794 \pm 0.019$
& $0.974 \pm 0.002$ \\

\rowcolor{candidate160row}
KetGPT \#160 & 0.70
& $36.12 \pm 3.86$ & 8398
& $\mathbf{81.78 \pm 1.23}$
& $\mathbf{0.819 \pm 0.017}$
& $\mathbf{0.820 \pm 0.013}$
& $\mathbf{0.816 \pm 0.017}$
& $\mathbf{0.977 \pm 0.002}$ \\

\addlinespace[2pt]

Classical & 0.80
& $50.4 \pm 4.8$ & 8460
& $79.10 \pm 1.42$
& $0.791 \pm 0.022$
& $0.793 \pm 0.014$
& $0.784 \pm 0.022$
& $0.972 \pm 0.003$ \\

\rowcolor{candidate160row}
KetGPT \#160 & 0.80
& $50.88 \pm 2.66$ & 8398
& $\mathbf{81.42 \pm 1.39}$
& $\mathbf{0.817 \pm 0.019}$
& $\mathbf{0.817 \pm 0.015}$
& $\mathbf{0.813 \pm 0.019}$
& $\mathbf{0.977 \pm 0.003}$ \\

\addlinespace[2pt]

Classical & 0.90
& $67.9 \pm 4.0$ & 8460
& $78.64 \pm 1.57$
& $0.787 \pm 0.024$
& $0.789 \pm 0.016$
& $0.779 \pm 0.024$
& $0.970 \pm 0.005$ \\

\rowcolor{candidate160row}
KetGPT \#160 & 0.90
& $66.34 \pm 2.61$ & 8398
& $\mathbf{81.44 \pm 1.39}$
& $\mathbf{0.818 \pm 0.019}$
& $\mathbf{0.817 \pm 0.015}$
& $\mathbf{0.814 \pm 0.018}$
& $\mathbf{0.977 \pm 0.003}$ \\

\bottomrule
\end{tabular}%
}
\end{table*}
\subsection{Low-Confidence Utility and Adaptive Routing}

We next examine whether the quantum fallback improves predictions on samples identified as uncertain by the primary classifier. On the low-confidence subset, Candidate \#160 improves accuracy over the matched MLP by $6.31$, $5.38$, and $3.73$ percentage points at $\tau=0.70$, $0.80$, and $0.90$, respectively.

At the validation-selected operating point $\tau=0.70$, Adaptive KetGPT \#160 reaches $81.78\%$ accuracy and $0.816$ F1, compared with $80.02\%$ accuracy and $0.794$ F1 for the adaptive classical baseline, while routing $36.12\%$ of test samples to the quantum branch.

\subsection{Resource Efficiency and Circuit Ablation}

Selective routing reduces quantum execution by invoking the quantum branch only for uncertain samples. With 100 shots per quantum evaluation, Adaptive KetGPT \#160 at $\tau=0.70$ routes $36.12\%$ of Fashion-MNIST samples, reducing the average quantum cost to $36.12$ shots per sample while reaching $81.78\%$ accuracy.

\begin{table}[h]
\centering
\caption{Resource ablation for 8-qubit KetGPT candidates on Fashion-MNIST. Results are mean $\pm$ sample standard deviation over five seeds.}
\label{tab:ketgpt_resource_five_seeds}
\small
\setlength{\tabcolsep}{6pt}
\renewcommand{\arraystretch}{1.05}
\begin{tabular}{lcccc}
\toprule
\textbf{ID} &
\textbf{Q. Params} &
\textbf{Gates} &
\textbf{Acc. (\%)} &
\textbf{F1} \\
\midrule
\#22  & 25 & 126 & $72.06 \pm 3.84$ & $0.706 \pm 0.049$ \\
\textbf{\#160} & 10 & 9 & $\mathbf{80.18 \pm 1.18}$ & $\mathbf{0.800 \pm 0.013}$ \\
\#180 & 58 & 49 & $76.72 \pm 2.02$ & $0.756 \pm 0.027$ \\
\bottomrule
\end{tabular}
\end{table}

This ablation uses a separate KetGPT candidate run with the lower parameter bound relaxed to include compact circuits. Candidate \#160 achieves the highest accuracy with only 10 quantum parameters and 9 gates, showing that higher circuit resource counts do not yield better performance in the evaluated candidates.

\subsection{Cross-Dataset Generalization}
\begin{table}[h]
\centering
\caption{Cross-dataset evaluation of KetGPT Candidate \#160 at a fixed routing threshold $\tau=0.90$. Results are mean $\pm$ sample standard deviation over five seeds. Routed denotes the percentage of test samples sent to the fallback branch.}
\label{tab:generalization_compact}
\small
\setlength{\tabcolsep}{5pt}
\renewcommand{\arraystretch}{1.10}
\resizebox{\columnwidth}{!}{%
\begin{tabular}{llcc}
\toprule
\textbf{Dataset} &
\textbf{Setting} &
\textbf{Routed (\%)} &
\textbf{Acc. (\%)} \\
\midrule

CIFAR-10
& Low-conf. MLP
& $63.5 \pm 16.1$
& $67.9 \pm 6.7$ \\

\rowcolor{candidate160row}
CIFAR-10
& Low-conf. KetGPT \#160
& $63.5 \pm 16.1$
& $72.2 \pm 4.8$ \\

CIFAR-10
& Adaptive Classical
& $63.5 \pm 16.1$
& $79.6 \pm 3.1$ \\

\rowcolor{candidate160row}
CIFAR-10
& Adaptive KetGPT \#160
& $63.5 \pm 16.1$
& $\mathbf{82.2 \pm 0.9}$ \\

\midrule

KMNIST
& Low-conf. MLP
& $88.0 \pm 7.9$
& $59.9 \pm 1.6$ \\

\rowcolor{candidate160row}
KMNIST
& Low-conf. KetGPT \#160
& $88.0 \pm 7.9$
& $62.2 \pm 1.4$ \\

KMNIST
& Adaptive Classical
& $88.0 \pm 7.9$
& $64.6 \pm 2.5$ \\

\rowcolor{candidate160row}
KMNIST
& Adaptive KetGPT \#160
& $88.0 \pm 7.9$
& $\mathbf{66.5 \pm 3.8}$ \\

\midrule

SVHN
& Low-conf. MLP
& $97.4 \pm 1.6$
& $44.5 \pm 1.9$ \\

\rowcolor{candidate160row}
SVHN
& Low-conf. KetGPT \#160
& $97.4 \pm 1.6$
& $45.7 \pm 3.7$ \\

SVHN
& Adaptive Classical
& $97.4 \pm 1.6$
& $45.9 \pm 1.8$ \\

\rowcolor{candidate160row}
SVHN
& Adaptive KetGPT \#160
& $97.4 \pm 1.6$
& $\mathbf{47.0 \pm 4.3}$ \\

\bottomrule
\end{tabular}%
}
\end{table}
We further evaluate whether the behavior observed on Fashion-MNIST is preserved across different image-classification datasets. Candidate \#160 is evaluated on CIFAR-10, KMNIST, and SVHN using a fixed routing threshold of $\tau=0.90$, without dataset-specific adjustment of the quantum fallback. The results in Table~\ref{tab:generalization_compact} show that KetGPT \#160 improves low-confidence accuracy over the MLP fallback on all three datasets. The corresponding adaptive system also improves over the adaptive classical baseline, with accuracy increasing from $79.6\%$ to $82.2\%$ on CIFAR-10, from $64.6\%$ to $66.5\%$ on KMNIST, and from $45.9\%$ to $47.0\%$ on SVHN. The routed fraction varies substantially across datasets, reaching $63.5\%$, $88.0\%$, and $97.4\%$, respectively, indicating that the amount of quantum invocation is determined by the confidence distribution of the primary classifier.

\section{Conclusion}
This paper introduced QSTAR, which uses confidence-based routing to invoke a searched quantum
fallback only for uncertain samples. Candidate \#160 combines compact
circuit structure with improved low-confidence and adaptive performance,
reaching $81.78\%$ accuracy while routing $36.12\%$ of Fashion-MNIST
samples. Consistent gains on CIFAR-10, KMNIST, and SVHN indicate that
selective quantum invocation generalizes beyond the primary dataset,
although its execution rate depends strongly on the classical
confidence distribution.



\section*{Acknowledgment}
 This work was supported in part by the NYUAD Center for Quantum and Topological Systems (CQTS), funded by Tamkeen under the NYUAD Research Institute grant CG008. This research was carried out on the High Performance Computing resources at New York University Abu Dhabi.

\section*{Generative AI Use Disclosure}

During the preparation of this work, the authors used Generative AI tools (specifically ChatGPT and Grammarly) for language editing, text refinement, and visual refinement of the methodology figure. The authors reviewed and edited all generated or refined content as needed and take full responsibility for the publication’s content.

\bibliographystyle{IEEEbib}
\bibliography{strings,refs}

@misc{he2015deepresiduallearningimage,
      title={Deep Residual Learning for Image Recognition}, 
      author={Kaiming He and Xiangyu Zhang and Shaoqing Ren and Jian Sun},
      year={2015},
      eprint={1512.03385},
      archivePrefix={arXiv},
      primaryClass={cs.CV},
      url={https://arxiv.org/abs/1512.03385}, 
}

@article{rehman2026fairbenchmark,
  title   = {Towards Fair Benchmarking of Quantum Transfer Learning for Visual Classification},
  author  = {Innan, Nouhaila and Rehman, Saim and Shafique, Muhammad},
  journal = {arXiv preprint arXiv:2605.19417},
  year    = {2026},
  eprint  = {2605.19417},
  archivePrefix = {arXiv},
  primaryClass = {cs.LG}
}

@article{mogalapalli2022classical,
  author    = {Harsha Mogalapalli and M. Abhuri and Nithya and S. K. V. Bendreddi},
  title     = {Classical-Quantum Transfer Learning for Image Classification},
  journal   = {SN Computer Science},
  volume2   = {3},
  number2    = {1},
  pages2     = {202},
  year      = {2022},
  doi2      = {10.1007/s42979-022-01074-3}
}

@article{kim2023classical,
  author    = {Juhyeon Kim and Joonsuk Huh and Daniel K. Park},
  title     = {Classical-to-Quantum Convolutional Neural Network Transfer Learning},
  journal   = {Neurocomputing},
  volume    = {555},
  pages2     = {126643},
  year      = {2023},
  doi2       = {10.1016/j.neucom.2023.126643}
}

@inproceedings{apak2024ketgpt,
  author    = {Boran Apak and others},
    author2    = {Boran Apak and Medina Bandi{\'c} and Aritra Sarkar and Sebastian Feld},
  title     = {KetGPT: Dataset Augmentation of Quantum Circuits using Transformers},
  booktitle = {International Conference on Computational Science (ICCS)},
  series    = {Lecture Notes in Computer Science},
  publisher = {Springer},
  year      = {2024},
  doi2      = {10.1007/978-3-031-63778-0_17}
}

@article{xiao2017fashion,
  author  = {Han Xiao and Kashif Rasul and Roland Vollgraf},
  title   = {Fashion-MNIST: A Novel Image Dataset for Benchmarking Machine Learning Algorithms},
  journal = {arXiv preprint arXiv:1708.07747},
  year    = {2017}
}

@article{mari2020transfer,
  title={Transfer learning in hybrid classical-quantum neural networks},
  author={Mari, Andrea and others},
    author2={Mari, Andrea and Bromley, Thomas R and Izaac, Josh and Schuld, Maria and Killoran, Nathan},
  journal={Quantum},
  volume1={4},
  pages2={340},
  year={2020},
  publisher1={Verein zur F{\"o}rderung des Open Access Publizierens in den Quantenwissenschaften}
}

@article{Schuld_2019,
   title={Evaluating analytic gradients on quantum hardware},
   volume={99},
   ISSN={2469-9934},
   url={http://dx.doi.org/10.1103/PhysRevA.99.032331},
   DOI={10.1103/physreva.99.032331},
   number={3},
   journal={Physical Review A},
   publisher={American Physical Society (APS)},
   author={Schuld, Maria and Bergholm, Ville and Gogolin, Christian and Izaac, Josh and Killoran, Nathan},
   year={2019},
   month=Mar }

@article{Wierichs_2022,
   title={General parameter-shift rules for quantum gradients},
   volume={6},
   ISSN={2521-327X},
   url={http://dx.doi.org/10.22331/q-2022-03-30-677},
   DOI={10.22331/q-2022-03-30-677},
   journal={Quantum},
   publisher={Verein zur Forderung des Open Access Publizierens in den Quantenwissenschaften},
   author={Wierichs, David and Izaac, Josh and Wang, Cody and Lin, Cedric Yen-Yu},
   year={2022},
   month=Mar, pages={677} }

@misc{geifman2017selectiveclassificationdeepneural,
      title={Selective Classification for Deep Neural Networks}, 
      author={Yonatan Geifman and Ran El-Yaniv},
      year={2017},
      eprint={1705.08500},
      archivePrefix={arXiv},
      primaryClass={cs.LG},
      url={https://arxiv.org/abs/1705.08500}, 
}

\end{document}